\documentclass{crc-book}
\usepackage{amsfonts, amsmath, amssymb, amsthm}

\usepackage[pdftex,
unicode=true,
]{hyperref}

\usepackage{cmap}

\usepackage{geometry,multicol}%[centering, margin=20mm]
\newdimen\stockheight
\newdimen\stockwidth
\stockwidth\paperwidth
\theoremstyle{definition}
\newtheorem{define}{Definition}
\newtheorem{Exmpl}{Example}
\newtheorem{rem}{Remark}

\DeclareMathOperator{\Lin}{Lin}
\DeclareMathOperator{\sym}{sym}
\DeclareMathOperator{\id}{id}

\newcommand{\sD}{\mathcal{D}}
\newcommand{\BBN}{\mathbb{N}}
\newcommand{\BBC}{\mathbb{C}}
\newcommand{\cR}{\mathcal{R}}

\begin{document}

%  Headings
%
\renewcommand{\evenhead}{A.~V.~Kiselev, A.~O.~Krutov, T.~Wolf}
\renewcommand{\oddhead}{Computing symmetries and recursion operators using SsTools}

%  Titlepage
%
\thispagestyle{empty}

\Name{Computing symmetries and recursion operators of evolutionary super-\/systems using the SsTools environment}

\Author{Arthemy V. Kiselev~$^a$ and Andrey O. Krutov~$^b$ and Thomas Wolf~$^c$}

\Address{$^a$ Johann Bernoulli Institute for Mathematics and Computer Science,
  University of Groningen,
  P.O.Box~407, 9700\,AK Groningen, The Netherlands.\\[10pt]
  $^b$ Independent University of Moscow,
  Bolshoj Vlasyevskij Pereulok 11,
  119002, Moscow, Russia.\\[10pt]
  $^c$ Department of Mathematics and Statistics,
  Brock University, 500 Glenridge Avenue, St.Catharines, 
  Ontario, Canada L2S 3A1
}

\begin{abstract}\noindent
At a \emph{very} informal but practically convenient level, we discuss
the step-\/by-\/step computation of nonlocal recursions for symmetry
algebras of nonlinear coupled boson-\/fermion $N=1$ supersymmetric
systems by using the \textsc{SsTools} environment.
\end{abstract}

The principle of symmetry plays an important role in modern
mathematical physics. The differential equations that constitute
integrable models practically always admit symmetry
transformations. The presence of symmetry transformation in a system
yields two types of explicit solutions: those which are invariant
under a transformation (sub)group and the solutions obtained by
propagating a know solution by the same group.  The recursion operator
is a (pseudo)differential operator which maps symmetries of a given
system to into symmetries of the same system. The recursion operators
allow to obtain new symmetries for a given seed symmetry.

It is common for important equations of mathematical physics 
not to have local recursion operators other than the
identity $\id\colon \varphi\mapsto \varphi$. Instead,
they often admit nonlocal recursions which involve integrations such
as taking the inverse of the total derivative~$D_x$ with respect to
the independent variable~$x$. To describe such nonlocal structures we
use the approach of nonlocalities. By nonlocalities we mean an
extension of the initial system by new fields such that the initial
fields are differential consequences of the new ones. In the case of
recursion operators such fields often arise for conservation laws. We
refer to a recursion operator for the Korteweg--\/de Vries equations
as a motivating example of a nonlocal recursion operator, see
Example~\ref{EXkdvrec} on page~\pageref{EXkdvrec}.

The supersymmetric integrable systems, i.e.\ systems involving
commuting (bosonic, or even) and anticommuting (fermionic, or odd)
independent variables and/or unknown functions, have found remarkable
applications in modern mathematical physics (for example supergravity
models, perturbed conformal field
theory~\cite{KupershmidtMathieu1989}; we refer to~\cite{Berezin,QFS}
for a general overview). When dealing with supersymmetric models of
theoretical physics, it is often hard to predict whether a certain
mathematical approximation will be truly integrable or not. Therefore
we apply the symbolic computation to exhibit necessary integrability
features. In what follows we restrict ourselves to the case of $N=1$
(where $N$ refers to the number of odd anticommuting independent variables
$\theta_i$). Nevertheless, the techniques and computer programs
described below could be easily applied to the case of
arbitrary~$N$. Usually, the $N$ is not bigger than~8, see for
example~\cite{KrivonosPashnevPopowicz1998}
and~\cite{DelducGallotIvanov1997}. It is an interesting open problem
to establish criteria that set a limit on $N$ in ``$N$-extended''
supersymmetric equations of mathematical physics.

The latest version of \textsc{SsTools} can be found
at~\cite{SsToolsURL}, see also~\cite{SsTools}.  We refer
to~\cite{Berezin,QFS} and~\cite{BVV,GDE2012,GradNonRem,KK2000,Olver}
for reviews of the geometry and supergeometry of partial differential
equations. We refer to~\cite{SymbolicBook} for an overview of other
software that could be used for similar computational tasks.

\section{Notation and definitions}

We fix notation first.
% \marginpar{Step~1.}
Let $x$ be the independent
variable, $u=(u^1,\ldots,u^m)$ denote the unknown functions
irrespective of their (anti-)\/commutation properties, and
$u_{k}=\{u^j_{k}|u^j_{k} = \partial^{k}u^j/\partial x^{k}\}$ denote
the partial derivatives of $u^j$ of order~$k\in\mathbb{N}$.
%\subsection{Superfields and superderivative.}  
We extend the
independent variable~$x$ by the pair $(x,\theta)$, where $\theta$ is
the Grassmann variable such that $\theta^2=0$. The superderivative is
defined as 
\[
\sD=\partial_{\theta}+\theta\partial_x.
\]
Its square power is the spatial derivative, $\sD^2=\partial_x$. 

Fields $u(x,t)$ now become $N=1$ superfields $u(x,t;\theta)$. Provided that $\theta^2=0$, they have a very
simple Taylor expansions in $\theta$:
\[
u(x,t;\theta)=u^0(x, t)+\theta u^1(x, t),
\]
here $u^0$ has the same parity as $u$ and $u^1$ has the opposite
parity. The bosonic fields (those commuting with everything) are
denoted by~$b(x,t;\theta)$, and the fermionic fields, which
anti-\/commute between themselves and with $\theta$, will be denoted
by~$f(x,t;\theta)$.  Further, we write $u_{k/2}=\sD^k u$ for the $k^{\rm th}$
order super-\/derivative of~$u$.  Note that the super-\/derivatives
$\sD^{2k+1}b$ are fermionic and $\sD^{2k+1}f$ are bosonic for
any~$k\in\mathbb{N}$.

Computer input will be shown in text font, for example,
\texttt{f(i)} for $f^i$,
\texttt{b(j)} for $b^j$,
\texttt{df(b(j),x)} for the derivative of~$b^j$ with respect to~$x$,
\texttt{d(1,f(i))} for the superderivative~$\sD_{\theta^1}f^i$ written as
$\sD f^i$, because we will have only one $\theta$ and one $\sD$.

Let $k,r<\infty$ be fixed integers. Suppose that $F^{\alpha}(x,t,u,\sD(u),\ldots,u_{k},u_t)$ is a smooth
function for any integer $\alpha\leq r$.
%\begin{define}[Differential equations]
In what follows, we consider systems of differential equations,
\begin{equation}\label{GenDE}
  \left\{F^{\alpha}(x,u,\sD(u),u_1,\sD(u_1),\ldots,u_{k},u_t)=0\right\},\quad \alpha = 1,\ldots,r,
\end{equation}
of order~$k$ and, especially, the autonomous translation-\/invariant
evolutionary systems,
\[
  \mathcal{E}=\left\{F^\alpha=u^\alpha_t-\Phi^\alpha(u,\ldots, u_k) = 0 \right\},
  \quad \alpha=1,\ldots,r,
\]
which are resolved w.r.t.\ the time derivatives, and the systems
obtained by extending the evolutionary systems with some further
differential relations upon $u$'s and their (super-)\/derivatives.
%\end{define}

\subsection*{The weight technique}
Physically meaningful equations have often several symmetries, among them
one or more scaling symmetries.
% \marginpar{Step~2.}
Suppose that to each super-\/field
$u^j$ and to the derivative~$\partial_t$ one can assign a real
number (the weight), which is denoted by $[A]$ for any object~$A$.  By
definition, $[\partial_x]\equiv1$. The weight of a product of two
objects is the sum of the weights of the factors, whence
$[\sD]=\frac{1}{2}$.  The weight of any nonzero constant equals $0$,
but the weight of a zero-valued constant can be arbitrary.

From now on, we consider differential equations~\eqref{GenDE} with
differential\/-\/polynomials $F^\alpha$ that admit the introduction of
weights for all variables and derivations such that the weights of all
monomials in each equation $F^\alpha=0$ coincide.  These equations
are scaling\/-\/invariant, or homogeneous.

\begin{Exmpl}
Consider the Burgers equation
\begin{equation}
\label{bBURG}
  b_t = b_{xx} - 2bb_x, \quad b=b(x,t).
\end{equation}
The weights are uniquely defined,
\[
[b]=1, \quad [\partial_t]=2 \Leftrightarrow [t]=-2, \quad [\partial_x]\equiv1.
\]
Indeed, equation~(\ref{bBURG}) is homogeneous w.r.t.\ these weights,
\[
 [b_t] = 1 + 2 = [b_{xx}] = 1 + 1 +1 = [2bb_x] = 0 + 1 + 1 + 1=  3
\]
and, clearly, this is the only way to choose the weights.
\end{Exmpl}

A system of differential equations could be homogeneous w.r.t.\ to different weight systems.
For a given system of differential equations these weight systems 
can be found by using the \texttt{FindSSWeight} function from \textsc{SsTools}:

\texttt{FindSSWeights(N,nf,nb,exli,zerowei,verbose)}\\
where
\begin{description} \itemsep-0.25em  
\item[\texttt{N}] \ldots~the number of superfields $\theta_i$;
\item[\texttt{nf}] \ldots~number of fermion fields \texttt{f(1), f(2), \ldots, f(nf)};
\item[\texttt{nb}] \ldots~number of boson   fields \texttt{b(1), b(2), \ldots, b(nb)};
\item[\texttt{exli}] \ldots~list of equations or expressions;
\item[\texttt{zerowei}] \ldots~list of constants or other kernels that should have zero weight;
\item[\texttt{verbose}] \ldots~(\texttt{=t}(true)/\texttt{nil}(false)) whether detailed comments shall be made.
\end{description}

The program returns a list of homogeneities, each homogeneity being a list~\texttt{fh, bh, hi} of
\begin{description}  \itemsep-0.25em  
\item[\texttt{fh}] \ldots~a list of the weights of \texttt{f(1),b(2), \ldots, f(nf)};
\item[\texttt{bh}] \ldots~a list of the weights of \texttt{b(1),b(2), \ldots, b(nb)};
\item[\texttt{hi}] \ldots~a list of the weights of equations/expressions in the input. 
\end{description}
Weights are scaled such that weight of $[\partial_x]$ is 2, i.e.\ the weight of any $\sD$ is 1.
So the computer weights will be twice the ``usual'' weight.
Input expressions can be in field form or coordinate form. 

\begin{Exmpl}
  Consider the nonlinear Schr\"odinger equation
  \begin{subequations}\label{eqSchr}
    \begin{align*}
      b^1_t = {}&   b^1_{xx} + 2 (b^1)^2b^2,& b^1={}&b^1(x,t),\\
      b^2_t = {}& - b^2_{xx} - 2 (b^2)^2b^1,& b^2={}&b^2(x,t).
    \end{align*}
  \end{subequations}
  We compute all possible weight systems of this system of equations:
\begin{verbatim}
FindSSWeights(0,0,2,{df(b(1),t) =   df(b(1),x,2) + 2*b(1)**2*b(2),
                     df(b(2),t) = - df(b(2),x,2) - 2*b(2)**2*b(1) },
              {},t)$
\end{verbatim}  
The output contains
\begin{verbatim}
This system has the following homogeneities:
W[t] = -4
W[b(2)] =  - arbcomplex(1) + 4
W[b(1)] = arbcomplex(1)
W[x] = -2
\end{verbatim}
which gives us the following family of weight systems for~\eqref{eqSchr}
\[
  [\partial_t] = 2,\quad [b^2] = - c + 4,\quad [b^1] = c,\quad [\partial_x] = 1,
\]
where $c$ is an arbitrary constant.

\end{Exmpl}

\section{Symmetries}
\begin{define}
An $l^{\rm th}$ order symmetry of an evolutionary system
$\mathcal{E}=\{u_t=\Phi\}$ is another autonomous evolutionary system
$\mathcal{E}'=\{u_s=\varphi(x,u,\sD(u),\ldots,u_l)\}$ upon
$u(s,t,x;\theta)$ such that a solution of the Cauchy problem for
$\mathcal{E}'$ propagates solutions of $\mathcal{E}$ to solutions
of~$\mathcal{E}$. A necessary and sufficient condition for a vector
%$\varphi=(F,B)^T$ 
$\varphi$ 
to be a symmetry of $\mathcal{E}$ is that solutions of the system satisfy
\begin{equation} \label{sycon}
D_t D_s u = \pm D_s D_t u
\end{equation}
where the minus sign applies if both $t$ and $s$ are fermionic.
Because (\ref{sycon}) is to be satisfied by solutions $u$ of the
system $\mathcal{E}$, i.e.\ $u_t$ is replaced by $\Phi$ giving the
so-called linearization $\Lin\mathcal{E}$ of the system $\mathcal{E}$:
\[  \Lin_\Phi(\varphi) := D_t \varphi = \pm D_s \Phi .\] 
This is a linear system for $\varphi$. If $s$ is a bosonic variable then
the symmetry is a system $\mathcal{E}'=\{f_s=F,\ b_s=B\}$ and if $s$ is
fermionic then the symmetry is $\mathcal{E}'=\{{f_s=B},\ b_s=F\}$.
\end{define}

The understanding of linearized systems $\Lin\mathcal{E}$ from a
computational viewpoint is as follows; we consider the differential
polynomial case since this is what \textsc{SsTools} can be 
applied to. Let us first consider the case of bosonic $s$.

Given a system $\mathcal{E}$ of super\/-\/equations, formally
assign the new `linearized` fields $F^i$ = \texttt{f(nf+i)} and $B^j$
= \texttt{b(nb+j)} to $f^i$ = \texttt{f(i)} and $b^j$ = \texttt{b(j)},
respectively, with $i=1$, $\ldots$, \texttt{nf} and $j=1$, $\ldots$,
\texttt{nb}. Pass through all equations, and whenever a power of a
derivative of a variable $f^i$ or $b^j$ is met, differentiate (in the
usual sense) this power with respect to its base, multiply the result
from the right by the same order derivative of $F^i$ or $B^j$,
respectively, and insert the product in the position where the power
of the derivative was met. Now proceed by the Leibniz rule. The final
result, when all equations in the system $\mathcal{E}$ are processed,
is the linearized system $\Lin\mathcal{E}$.

If $s$ is fermionic then proceed in the same way, except we get extra factors of $-1$:
\begin{itemize}
\item An overall factor $-1$ appears in (\ref{sycon}) if $t$ is fermionic 
      due to anticommuting $D_t$ and $D_s$.
\item When differentiating a factor in a product then a factor
      $-1$ appears for each fermionic factor to the left of the
      differentiated factor.
\item When changing the order of $D_s$ and $\sD$ in differentiating then a factor
      of $-1$ appears as well.
\end{itemize}

The second difference between bosonic and fermionic $s$ is the number of
new `linearized` fields $F^i, B^j$ that are introduced in the linearized equation.
For bosonic $s$ these are $F = \texttt{f(nf+1)} \ldots \texttt{f(nf+nf)}, \ \ \ 
B = \texttt{b(nb+1)} \ldots \texttt{b(nb+nb)}$ whereas for fermionic $s$ these are
$F = \texttt{f(nf+1)} \ldots \texttt{f(nf+nb)}, \ \ \ 
B = \texttt{b(nb+1)} \ldots \texttt{b(nb+nf)}$.

To summarize, the linearization is obtained by a complete differentiation
$D_s \Phi$ applying the Leibniz rule and chain rule (in place) and substituting
$u_s = \varphi$.

\begin{Exmpl}
The linearized counterpart of 
$b_t = b(\sD f )^2$ is $B_t = B(\sD f)^2 + 2b \sD f\sD F$
for bosonic symmetry parameter $s$ and
$F_t = F(\sD f)^2 - 2b \sD f\sD B$ for fermionic $\bar s$.
Likewise, for $b_t=b f f_x (Df)^2$ and parity-odd $\bar{s}$, the
linearization is $F_t = F f f_x (Df)^2 + b B f_x (Df)^2 - b f B_x
(Df)^2 - 2b f f_x Df DB$.
\end{Exmpl}

The scaling weights of the new fields are always set by $[F]=[f]$, $[B]=[b]$
for bosonic $s$ and  $[F]=[b]$, $[B]=[f]$ for fermionic $\bar s$.

The linearization $\Lin\mathcal{E}$
%\marginpar{Step~3.}
  for a system
of evolution equations $\mathcal{E}$ is obtained using the procedure
\texttt{linearize}:\\ \texttt{linearize(pdes, nf, nb, tpar, spar);}\\ where
\begin{description}\itemsep-0.25em  
 \item[\texttt{pdes}] \ldots~list of equations 
                      $b^i_t = \Phi^i_b,\quad f^j_t = \Phi^j_f$;
 \item[\texttt{nf}] \ldots~number of the fermion fields 
                    \texttt{f(1), f(2), \ldots, f(nf)};
 \item[\texttt{nb}] \ldots~number of the boson fields 
                    \texttt{b(1), b(2), \ldots, b(nb)};
 \item[\texttt{tpar}] \ldots~(=\texttt{t}(true)/\texttt{nil}(false)) whether $t$ is parity changing or not
 \item[\texttt{spar}] \ldots~(=\texttt{t}(true)/\texttt{nil}(false)) whether $s$ is parity changing or not
\end{description}

\begin{Exmpl}
The linearizations of the system with parity reversing time~$\bar{t}$,
\begin{equation}\label{EqTPar}
 f_{\bar t} = b^2 + \sD f, \quad b_{\bar t}=fb+\sD b,
\end{equation}
are obtained as follows,
\begin{verbatim}
depend {b(1),f(1)},x,t;
linearize({df(f(1),t)=b(1)**2+d(1,f(1)),
           df(b(1),t)=f(1)*b(1)+d(1,b(1))},1,1,t,nil); 
\end{verbatim}
for bosonic $s$, and the same call with \texttt{t} as last parameter 
instead of \texttt{nil} for fermionic $\bar s$.

The result is the new system involving twice as many variables as
the original equation. The linearization correspondence between the
fields is
\[f \mapsto F \quad (\texttt{f(1)}\mapsto\texttt{f(2)}), \qquad
  b \mapsto B \quad (\texttt{b(1)}\mapsto\texttt{b(2)}) \]
for bosonic $s$ and
\[f \mapsto B \quad (\texttt{f(1)}\mapsto\texttt{b(2)}), \qquad
  b \mapsto F \quad (\texttt{b(1)}\mapsto\texttt{f(2)}) \]
for fermionic ${\bar s}$.

The procedure to compute the linearization is the same for normal
times $t$ and for parity reversing times ${\bar t}$ 
except of a factor (-1) of the rhs's $B,F$ if both ${\bar t, \bar s}$
are fermionic (because of the anticommutativity of $D_{\bar t}, D_{\bar s}$
in that case).

The linearized system incorporates
\begin{itemize}
 \item  the initial system:\\
 \texttt{df(f(1),t) = b(1)**2 + d(1,f(1)),}\\
 \texttt{df(b(1),t) = f(1)*b(1) + d(1,b(1)),}

 \item and its linearizations:\\
 \texttt{df(f(2),t) = 2*b(2)*b(1) + d(1,f(2)),}\\ 
 \texttt{df(b(2),t) = d(1,b(2)) + f(2)*b(1) + f(1)*b(2)}\\
 for $s$; respectively,\\
 \texttt{df(b(2),t) = d(1,b(2)) - 2*f(2)*b(1),}\\
 \texttt{df(f(2),t) = - b(2)*b(1) + d(1,f(2)) - f(2)*f(1)}\\
%\end{verbatim}
 % \texttt{df(b(2),t) = 2*b(1)*f(2) - d(1,b(2)),}\\ 
 % \texttt{df(f(2),t) = - d(1,f(2)) + b(2)*b(1) - f(1)*f(2)}\\
 for $\bar s$.
\end{itemize}
\end{Exmpl}

One does not need to compute the linearizations in
order to obtain a symmetry of a differential equation. However, the
explicit computation of the linearizations will be required for
finding the recursions, which are ``symmetries of symmetries.''

For computing symmetries of any system $\mathcal{E}$, use the
procedure \texttt{ssym} with the call\\
\texttt{ssym(N,tw,sw,afwlist,abwlist,eqnlist,fl,inelist,flags);}
\begin{description}\itemsep-0.25em  
 \item[\texttt{N}] $\ldots$~the number of superfields $\theta_i$;
 \item[\texttt{tw}] $\ldots$~$2 \times$the weight of $\partial_t$;
 \item[\texttt{sw}] $\ldots$~$2 \times$the weight of $\partial_s$;
 \item[\texttt{afwlist}] $\ldots$~list of $2 \times$weights of the fermion fields 
                         \texttt{f(1),f(2),$\ldots$,f(nf)};
 \item[\texttt{abwlist}] $\ldots$~list of $2 \times$weights of the boson \ \ fields 
                         \texttt{b(1),b(2),$\ldots$,b(nb)};
 \item[\texttt{eqnlist}] $\ldots$~list of extra conditions on the 
                         undetermined coefficients;
 \item[\texttt{fl}] $\ldots$~extra unknowns in \texttt{eqnlist} to be determined;
 \item[\texttt{inelist}] $\ldots$~a list, each element of it is a non-zero
                         expression or a list with at least one of 
                         its elements being non-zero;
 \item[\texttt{flags}] $\ldots$~list of flags:\\
             \ \ \ \ \ \texttt{init}: only initialization of global data,\\
             \ \ \ \ \ \texttt{zerocoeff}: all coefficients = 0 which do 
                                           not appear in inelist,\\
             \ \ \ \ \ \texttt{tpar}: if the time variable \(t\) changes parity,\\
             \ \ \ \ \ \texttt{spar}: if the symmetry variable \(s\) changes parity,\\
             \ \ \ \ \ \texttt{lin}: if symmetries of a linearization are 
                                     to be computed,\\
             \ \ \ \ \ \texttt{filter}: if a symmetry should satisfy homogeneity 
                                        weights defined in \texttt{hom\_wei} 
\end{description}
Note that the computer representation of the weights is twice the standard
notation; thus we avoid  half\/-\/integers values for convenience.
For more details on other flags run the command \texttt{sshelp()}.

\section{Recursions}
\begin{define}
A recursion operator for the symmetry algebra of an evolutionary system 
\[
\mathcal{E}=\{f_t^i=\Phi^i_f,\ b^i_t=\Phi^j_b\}, \ \ i=1,\ldots,n_f,\ \ j=1,\ldots,n_b
\]
is the vector expression 
\[
\mathcal{R}=(\mathcal{R}^1_f(\varphi), \ldots, \mathcal{R}^{n_f}_f(\varphi), 
             \mathcal{R}^1_b(\varphi), \ldots, \mathcal{R}^{n_b}_b(\varphi))^{T},
\]
which is linear w.r.t.\ the new fields $\varphi$, which for bosonic $s$ are
$\varphi=(F^1,\ldots,F^{n_f},B^1,$ $\ldots,B^{n_b})^T$ and for fermionic $\bar s$ are 
$\varphi=(B^1,\ldots,B^{n_f},F^1,\ldots,F^{n_b})^T$,
and their derivatives and which is a
(right-\/hand side of a) symmetry of $\mathcal{E}$ whenever
\[\varphi=(F^1, \ldots, F^{n_f}, B^1, \ldots, B^{n_b})^{T} \ \ \ \mbox{for $s$, or}\] 
\[\varphi=(B^1, \ldots, B^{n_f}, F^1, \ldots, F^{n_b})^{T} \ \ \ \mbox{for $\bar s$}\ \ \ \ \] 
is a symmetry of $\mathcal{E}$. In other words,
$\mathcal{R}\colon\sym\mathcal{E}\to\sym\mathcal{E}$ is a linear
operator that generates a symmetry of $\mathcal{E}$ when applied to
a symmetry $\varphi \in \sym \mathcal{E}$.

The weight $[\mathcal{R}]$ of recursion $\mathcal{R}$ is the difference 
$[\mathcal{R}(\varphi)]-[\varphi]$ of the weights of the (time derivatives 
$\partial_{s'}$, $\partial_s$ of $u$, i.e.\ of the) 
resulting and the initial symmetries, here $\varphi\in\sym\mathcal{E}$.
\end{define}

Different recursions can have the same weight. Experiments show that
in this case operators may have different properties, e.g., a majority
of them is nilpotent, several zero\/-\/order operators act through
multiplication by a differential\/-\/functional expression and do not
increase the differential orders of the flows, and only few recursions
construct higher\/-\/order symmetries and reveal the integrability.

If $[\mathcal{R}]$ is the weight of a recursion~$\mathcal{R}$, then,
clearly, at least one recursion is found with weight
$2\times[\mathcal{R}]$. Indeed, this is $\mathcal{R}^2$. At the same
time, other recursions may appear with weight 
$\mathbb{N}\times[\mathcal{R}]$. If $\mathcal{R}$ is nonlocal (see
below), then its powers are also nonlocal, but it remains a very
delicate matter to predict the form of their nonlocalities, and strong
theoretical assertions can be formulated for some particular
integrable system.

\begin{rem}\label{remW}
The derivation of the weight of a recursion is constructive in the
following sense. To attempt finding a recursion for an
evolutionary system, it is beneficial to know already many symmetries
$\varphi_{s_i}$ of different weights $[s_i]$. Then
one tries first the weights $[\mathcal{R}]\mathrel{{:}{=}}[s_i]-[s_j]$
for various $i,j$. However, the recursions obtained this way can be
nilpotent, i.e.\ $\mathcal{R}^q(\varphi)\equiv0$ for some~$q$ and
any~$\varphi$. More promising are weights for which there
exist $i,j,k$ with $[\mathcal{R}]=[s_i]-[s_j]=[s_j]-[s_k]$
and $s_i, s_j, s_k$ being elements of the infinite hierarchies
of symmetries with low weights. Still, the actual weights of unknown
recursions can turn out to be larger than the weight differences of the
lowest order symmetries.

% Second, the growing weights of the symmetries may correspond to the
% flows propagated by a recurrence relation \cite{Kiev2005} and not by a
% recursion operator, which in fact means the functional arbitrariness
% with respect to some expression in a set of commuting flows. 
% This does not seem interesting in the theory of integrable systems.

Finally, non-trivial recursions may only appear in nonlocal settings.
We discuss this in section~\ref{SecNonlocal}.
\end{rem}

The crucial point is that $\mathcal{R}$~is a symmetry of the
linearized system $\Lin\mathcal{E}$.  The original system
$\mathcal{E}$ is only used for substitutions. Hence we use
\texttt{ssym} for finding recursions of the linearizations, which are
previously calculated by \texttt{linearize}. %\marginpar{Step~4.}

\begin{Exmpl}
Let us construct a recursion for equation~(\ref{bBURG}). 
We obtain the linearization using the procedure \texttt{linearize},
\begin{verbatim}
linearize({df(b(1),t)=df(b(1),x,2)-2*b(1)*df(b(1),x)},0,1,nil,nil);
\end{verbatim}
The new system depends on the two fields $b=\texttt{b(1)}$ and 
$B=\texttt{b(2)}$.
\begin{verbatim}
df(b(2),t) = -2*b(2)*df(b(1),x)-2*b(1)*df(b(2),x)+df(b(2),x,2),
df(b(1),t) = df(b(1),x,2)-2*b(1)*df(b(1),x).
\end{verbatim}
The recursion of weight $0$ is obtained as follows (thus 
$\texttt{sw}=2[\mathcal{R}]=0$).
\begin{verbatim}
ssym(1, 4, 0, {}, {2, 2},
  {df(b(2),t)= -2*b(2)*df(b(1),x)-2*b(1)*df(b(2),x)+df(b(2),x,2),
   df(b(1),t)=> df(b(1),x,2)-2*b(1)*df(b(1),x)}, {}, {}, {lin}); 
\end{verbatim}
By writing~\texttt{df(b(1),t)=> ...} we require that the original system
is only used for substitutions.

The output contains\begin{verbatim}df(b(2),s)=b(2)
1 solution was found.
\end{verbatim}
This is the identity transformation
\[
 \varphi \mapsto \mathcal{R} = \text{id}\,(\varphi)\equiv\varphi;
\]
it maps symmetries to themselves.
Clearly, the identity is a recursion for any system\,!
\end{Exmpl}

\begin{Exmpl}[A recursion for the Korteweg\/--\/de Vries equation]
The KdV equation upon the bosonic field $u(x,t)$ is
\begin{equation}
\label{bKdV}
 u_t = - u_{xxx} + uu_x. 
\end{equation}
Equation~(\ref{bKdV}) is homogeneous w.r.t.\ the weights
\[
 [u] = 2,\quad [t] = -3,\quad [x] = -1.
\]
The linearization of~(\ref{bKdV}) is constructed using the procedure \texttt{linearize},
\begin{verbatim}
linearize({df(b(1),t)= -df(b(1),x,3) + b(1)*df(b(1),x)},0,1,nil,nil)
\end{verbatim}
Thus we obtain the new system that depends on the fields $u=\texttt{b(1)}$, $U=\texttt{b(2)}$:
\begin{verbatim}
df(b(1),t) = - df(b(1),x,3) + b(1)*df(b(1),x)
df(b(2),t) = b(2)*df(b(1),x) + b(1)*df(b(2),x) - df(b(2),x,3).
\end{verbatim}
As a first guess we are looking for the recursion operator of weight~$0$.
The recursion of weight $0$ (hence $\texttt{sw} = -2[\mathcal{R}] = 0$) is constructed as follows.
\begin{verbatim}
ssym(1, 6, 0, {}, {4, 4},
     {df(b(2),t) =  b(2)*df(b(1),x)+b(1)*df(b(2),x)-df(b(2),x,3),
      df(b(1),t) => -df(b(1),x,3)+b(1)*df(b(1),x)},{},{},{lin});
\end{verbatim}
The output contains
\begin{verbatim}
df(b(2),s)=b(2)
1 solution was found.
\end{verbatim}
Again, this operator $\mathcal{R}$ of weight $0$ is the identity,
\[
 \varphi \mapsto \mathcal{R}(\varphi) = \left(\text{id}\right)(\varphi).
\]
The well-\/known explanation for this result is that,
as a rule, one needs to introduce nonlocalities first and only then
obtains nontrivial recursions in the nonlocal setting.
\end{Exmpl}

\section{Nonlocalities}\label{SecNonlocal}
The nonlocal variables for $N=1$ super-\/systems are constructed by 
trivializing~\cite{Kiev2005,KK2000} conservation laws
\[
\partial_t(\text{density})\doteq\sD(\text{super\/-flux}),
\]
% which are conserved on these systems,
that is, in each case the above equality holds by virtue ($\doteq$) of
the system at hand and all possible differential consequences from it.  The standard procedure~\cite{KK2000}
suggests that every conserved current determines the new nonlocal variable, say~$v$, whose derivatives are set
to %\marginpar{Step~5.}
\begin{subequations}\label{EqNewNonlocal}
\begin{align}
v_t&=\text{super\/-\/flux},\quad \sD v=\text{density}\\
\intertext{if the time~$t$ preserves the parities and}
v_{\bar t}&=-\text{super\/-\/flux},\quad \sD v=\text{density}\\
\intertext{if the time~$\bar{t}$ is parity\/-\/reversing.
Note that in the classical case the nonlocality~$v$ can be specified through}
v_t&=\text{flux},\quad v_x=\text{density}
\end{align}
\end{subequations}
for the conservation law $\partial_t(\text{density})\doteq\partial_x(\text{flux})$.  Each nonlocality thus
makes the conserved current trivial because the cross derivatives of $v$ coincide in this case,
$[D_t,\sD](v)=0$, where $[\,,\,]$ stands for the commutator if $t$ is parity\/-\/preserving and for the
\emph{anti}commutator whenever the time~$\bar{t}$ is parity\/-\/reversing.  The new variables can be bosonic
or fermionic; the parities are immediately clear from the formulae for their derivatives.

Hence, starting with an equation~$\mathcal{E}$, one calculates several conserved %\marginpar{Step~6.}
currents for it and \emph{trivializes} them by introducing a \emph{layer} of nonlocalities whose derivatives
are still local differential functions. This way the number of fields is increased and the system is extended
by new substitution rules. Moreover, it may acquire new conserved currents that depend on the nonlocalities
and thus specify the second layer of nonlocal variables with nonlocal derivatives. At each step the number of
variables will increases by 2 compared with the previous layer (a new nonlocal variable plus the corresponding
linearized field). Clearly, the procedure is self-\/reproducing.

So, one keeps computing conserved currents and adding the layers of nonlinearities until an extended
system~$\smash{\tilde{\mathcal{E}}}$ is achieved such that its linearization~$\Lin\smash{\tilde{\mathcal{E}}}$
has a
 %(rigorously speaking \cite{JKKersten}, a shadow of a) 
symmetry~$\mathcal{R}$; this symmetry of $\Lin\smash{\tilde{\mathcal{E}}}$ is a recursion for the extended
system~$\smash{\tilde{\mathcal{E}}}$. %\marginpar{Success!}

  %ssconl   This is discovered by \SsTools\ as follows.
The calculation of conservation laws for
evolutionary super\/-\/systems with homogeneous polynomial
right\/-\/hand sides is performed by using the procedure
\verb"ssconl":
\begin{verbatim}
 ssconl(N,tw,mincw,maxcw,afwlist,abwlist,pdes);
\end{verbatim}
where
\begin{description}\itemsep-0.25em  
\item[\texttt{N}] \ldots~the number of superfields $\theta^i$;
\item[\texttt{tw}]  \ldots~$2\times$the weight $[\partial_t]$;
\item[\texttt{mincw}] \ldots~minimal weight of the conservation law;
\item[\texttt{maxcw}] \ldots~maximal weight of the conservation law;
\item[\texttt{afwlist}] \ldots~list of weights of the fermionic fields \texttt{f(1)},$\ldots$,\texttt{f(nf)};
\item[\texttt{abwlist}] \ldots~list of weights of the bosonic fields \texttt{b(1)},$\ldots$,\texttt{b(nb)};
\item[\texttt{pdes}]  \ldots~list of the equations for which a conservation law must be found.
\end{description}

% \smallskip
\noindent%
%The ansatz for the differential polynomial components of a conserved current is composed in full generality.
% (certainly, it has nothing to do with the axioms for the symmetry flows).
For positive weights of bosonic variables, the ansatz is fully determined through the weight
\texttt{mincw, \ldots, maxcw} of the conservation law.  If a boson weight is non-positive then the global variable
\verb'max_deg' must have a positive integer value which is the highest degree of such a variable or any of
its derivatives in any ansatz.  The conservation law condition leads to an algebraic system for the
undetermined coefficients, which is further solved automatically by \textsc{Crack}.

% Again, the global positive integer variable \verb"max_deg"
% determines the highest power of bosonic variables of nonpositive weights and all their (super-)\/derivatives
% in any ansatz.
% %The procedure \verb"ssconl" (and \verb"wgts" and \verb"linearize" as well)
% %is indifferent w.r.t.\ the presence of assignments \verb"=" and \verb"=>" 
% %in \verb"pdes".
% The fact that the current is conserved on a given system
% \verb"pdes" leads to the algebraic system,
% % for the undetermined coefficients,
%  which is further solved automatically by
% \textsc{Crack}. %~\cite{WolfCrack}. 
Having obtained a conserved current, one defines
the new bosonic or fermionic dependent variable (the nonlocality) using
the standard rules~\eqref{EqNewNonlocal}.
    %formulated in section~\ref{SecEqns}.

We illustrate the general scheme of fixing the derivatives of a nonlocal variable
by several examples. Further information on the \textsc{SsTools} environment is contained
in~\cite{SsTools} and the \texttt{sshelp()} function in \textsc{SsTools}. The algebraic structures that
describe the geometry of recursion operators for super-\/PDE are described in detail in~\cite{KK2000}. Some
more examples and their applications are also found in~\cite{Kiev2005}. 
%An example illustrating the run of \verb"ssconl" is given in
%the test output. %appendix~\ref{App}.

\begin{Exmpl}[A nonlocal recursion for the KdV equation]\label{EXkdvrec}
Consider the Korteweg\/--\/de Vries equation~\eqref{bKdV} again,
\[
\partial_t(u)\doteq\partial_x\Bigl(-u_{xx}+\frac12u^2\Bigr).
\]
We declare that the conserved density $u$ is the spatial derivative $w_x=u$ of a new nonlinear variable~$w$ and
the flux is its derivative w.r.t.\ the time, $w_t=-u_{xx}+\frac12u^2$. Then $w_{xt}=w_{tx}$ by virtue
of~\eqref{bKdV}.  Thus we introduce the bosonic nonlocality $w$ by trivializing the conserved current.  Let
us remember that
\begin{align*}
w_x &= u,\\
w_t &= - w_{xxx} + \frac{1}{2}w_x^2
\end{align*}
and the weight of $w$ is $[w]=1$ because $[w]+[\partial_x]=[u]=2$.

Next, we compute the linearization of equation~\eqref{bKdV} and of the relations that specify the new variable,
\begin{verbatim}
linearize({df(b(1),t)= -df(b(1),x,3) + b(1)*df(b(1),x),
           df(b(2),x)= b(1),
           df(b(2),t)= -df(b(2),x,3) + df(b(2),x)**2}, 0, 2);
\end{verbatim}
The linearization correspondence between the fields is
\[
 u \mapsto U \quad( \texttt{b(1)} \mapsto \texttt{b(3)}), \qquad w \mapsto W \quad (\texttt{b(2)} \mapsto \texttt{b(4)}).
\]
The linearized system is
\begin{verbatim}
df(b(3),t) = b(3)*df(b(1),x) + b(1)*df(b(3),x) - df(b(3),x,3),
df(b(4),x) = b(3),
df(b(4),t) = -df(b(4),x,3) + df(b(2),x)*df(b(4),x)
\end{verbatim}
In this nonlocal setting, we obtain the nonlocal recursion of weight $\texttt{sw}=2[\mathcal{R}]=4$ as follows,
\begin{verbatim}
ssym(1, 6, 4, {}, {4, 2, 4, 2},{
    df(b(3),t) = b(3)*df(b(1),x) + b(1)*df(b(3),x) - df(b(3),x,3),
    df(b(1),t) => b(1)*df(b(1),x) - df(b(1),x,3),
    df(b(2),x) => b(1),
    df(b(2),t) => -df(b(2),x,3) + 1/2 * df(b(2),x)**2,
    df(b(4),x) => b(3),
    df(b(4),t) => -df(b(4),x,3) + df(b(2),x)*df(b(4),x)
}, {}, {}, {lin});
\end{verbatim}
We recall here that only the linearized system should be written as equations, and all other relations,
including the nonlocalities, should be written as substitutions.

This yields the solution
\begin{verbatim}
 df(b(3),s) = -3*df(b(3),x,2) + 2*b(1)*b(3) + df(b(1),x)*b(4),
\end{verbatim}
which is the well-\/known nonlocal recursion operator for KdV,
\begin{equation}\label{reckdv}
 \varphi \mapsto \mathcal{R} = 
 \left( -3D_x^2 + 2u + u_x\cdot D_{x}^{-1}\right)(\varphi) .
\end{equation}
This recursion generates the hierarchy of \emph{local} symmetries starting from the translation
$\varphi_0=u_x$. The powers $\mathcal{R}^2$, $\mathcal{R}^3$, $\ldots$ of the recursion operator are also
nonlocal.
% but each of them requires the introduction of only \emph{one} nonlocality for~\eqref{bKdV} by
% trivializing \emph{higher} conservation laws from the hierarchy of the Korteweg\/--\/de Vries equation.
\end{Exmpl}

\begin{Exmpl}
Consider the Burgers equation~(\ref{bBURG}) and introduce the bosonic nonlocality $w$ of weight $[w]=[b]-[\partial_x]=0$ by trivializing the conserved current 
$\partial_t(b)\doteq\partial_x\left( b_x - b^2 \right)$.
We therefore, set 
\begin{equation}\label{pbBURG}
w_x = b,\quad w_t = w_{xx} - w_x^2.
\end{equation}
The linearization of the extended system is obtained through
\begin{verbatim}
linearize({df(b(1),t)=df(b(1),x,2) - 2*b(1)*df(b(1),x),
           df(b(2),x)=b(1),
           df(b(2),t)=df(b(2),x,2) - df(b(2),x)**2}, 0, 2);
\end{verbatim}
The correspondence between the bosonic fields is
\[
 b \mapsto B \quad( \texttt{b(1)} \mapsto \texttt{b(3)}), \quad w \mapsto W\quad ( \texttt{b(2)} \mapsto \texttt{b(4)}).
\]
The entire linearized system is~(\ref{bBURG}) and~(\ref{pbBURG})
together with the relations
\begin{verbatim}
df(b(3),t)= -2*b(3)*df(b(1),x)-2*b(1)*df(b(3),x)+df(b(3),x,2);
df(b(4),x)= b(3);
df(b(4),t)= df(b(4),x,2) - 2*df(b(4),x)*df(b(2),x).
\end{verbatim}
The difference between weights of the first-order and the second-order symmetries
is~$1$. Hence, the recursion operator could have weight~$1$, see Remark~\ref{remW}.
The nonlocal recursion of weight $1$, $\texttt{sw}=2[\mathcal{R}]=2$, is obtained by 
\begin{verbatim}
max_deg:=1;
ssym(1, 4, 2, {}, {2, 0, 2, 0},{
    df(b(3),t) = -2*b(3)*df(b(1),x)-2*b(1)*df(b(3),x)+df(b(3),x,2), 
    df(b(4),x) => b(3),
    df(b(4),t) => df(b(4),x,2)-2*df(b(4),x)*df(b(2),x),
    df(b(1),t) => -2*b(1)*df(b(1),x)+df(b(1),x,2),
    df(b(2),x) => b(1),
    df(b(2),t) => df(b(2),x,2)-df(b(2),x)**2}, {},{},{lin});
\end{verbatim}
We finally get the recursion
\begin{verbatim}
df(b(3),s) = -df(b(3),x) + b(1)*b(3) + df(b(1),x)*b(4),
\end{verbatim}
which is nonlocal,
\[
 \varphi \mapsto \mathcal{R} = \left(-D_x+b+b_x D_x^{-1} \right)(\varphi).
\]
\end{Exmpl}

\begin{Exmpl}
Consider the super\/-\/field representation~\cite{Kiev2005} of the Burgers equation, see~\eqref{bBURG},
\[
 f_t = \sD b, \quad b_t = \sD f + b^2;
\]
its weights are $|f|=|b|=\frac12$, $|\partial_x|=1$, and $|\partial_t|=\frac12$. 
 
We introduce the nonlocal bosonic field $w(x,t;\theta)$ of weight $[w]=[f]-[\sD]= 1-1=0$ such that
\[
 \sD w = -f, \quad w_t = -b.
\]
We get the linearized system by
\begin{verbatim}
linearize({df(f(1),t)= d(1, b(1)), 
           df(b(1),t)= df(f(1),x) + b(1)**2,
           d(1,b(2)) = -f(1), 
           df(b(2),t)= -b(1)}, 1, 2);
\end{verbatim}
For the linearization correspondence between the fields is $f\to F$, $b\to B$, $w\to W$, and we have
\[
  F_t = \sD B, \quad B_t=2bB +\sD F, \quad \sD W = -F, \quad W_t = -B,
\]
that is,
\begin{verbatim}
df(f(2),t)= d(1, b(3)),
df(b(3),t)= 2*b(1)*b(3)+d(1,f(2)),
d(1,b(4)) = -f(2),
df(b(4),t)= -b(3).
\end{verbatim}
In this setting, we obtain the nonlocal recursion of weight $\frac12$, $\texttt{sw}=2[\mathcal{R}]=1$: the input is
\begin{verbatim}
max_deg:=1;
ssym(1, 1, 1, {1, 1}, {1, 0, 1, 0},
     {df(f(2),t) = d(1, b(3)),
      df(b(3),t) = 2*b(3)*b(1) + d(1, f(2)),
      d(1, b(4)) => - f(2),
      df(b(4),t) => - b(3),
      df(f(1),t) => d(1, b(1)),
      df(b(1),t) => d(1, f(1)) + b(1)**2,
      d(1, b(2)) => -f(1),
      df(b(2),t) => -b(1)}, {}, {}, {lin});
\end{verbatim}
The recursion is
\begin{verbatim}
df(f(2),s)=d(1,b(3)) + d(1,b(1))*b(4) - f(2)*b(1),
df(b(3),s)=b(4)*b(1)**2 + b(3)*b(1) + d(1,f(2)) + d(1,f(1))*b(4),
\end{verbatim}
in other words,
\[ \left(\begin{array}{c} F \\ B \end{array} \right) \mapsto 
   \mathcal{R} \left( \begin{array}{c} F \\ B \end{array}\right) = 
   \left(
   \begin{array}{c}
    \sD B - \sD b \sD^{-1}F - Fb   \\ 
    - b^2\sD^{-1} F + Bb + \sD F - \sD f\sD^{-1}F
   \end{array} \right).
\]
\end{Exmpl}

\begin{Exmpl}
Consider the fifth order evolution superequation found by Tian and Liu (Case F in~\cite{TianLiu5ord}, see
also~\cite{GradNonRem,TianWang2016}):
\begin{equation}\label{eq5ordFsuper}
    f_t = f_{5x} + 10 (f_{xx}\sD f)_x + 5 (f_x\sD f_x)_x + 15f_x(\sD f)^2 + 15 f(\sD f_x)(\sD f).
\end{equation}
In what follows, we are considering this equations in components. Substitution $\xi + \theta u$ for $f$
in~\eqref{eq5ordFsuper}, for example, using \textsc{SsTools}, we obtain
\begin{subequations}\label{eq5ordF}
  \begin{align}
    u_t = {} & {} u_{5x} + 10 uu_{xxx} + 20u_xu_{xx} + 30 u^2u_x - 5\xi_{xxx}\xi_x 
               + 15u\xi_{xx}\xi + 15u_x\xi_x\xi,\label{eq5ordFBosonic}\\
    \xi_t={} & {} \xi_{5x} + 10u\xi_{xxx} + 15u_x\xi_{xx} + 5u_{xx}\xi_x + 15u^2\xi_x + 15uu_x\xi,
  \end{align}
\end{subequations}
where $u$ is a bosonic field and $\xi$ is a fermionic field.

Observe that the bosonic limit ($\xi:=0$) is the fifth order symmetry of Korteweg--\/de Vries
equation~\eqref{bKdV}. However, a direct computation shows that the equation~\eqref{eq5ordF} has local
symmetries of the orders $1 + 6k$ and $5+6k$, where $k\in\BBN$, and does not have any local symmetries of order
$3+6k$, where $k\in\BBN$. Therefore, the recursion operator for~\eqref{eq5ordF} should be at least of order
$6$. Let us also assume that the bosonic limit of the recursion operator for~\eqref{eq5ordF} is the third
power~$\cR^3$ of the recursion operator~\eqref{reckdv} for the Korteweg--\/de Vries equation.

It is easy to check that for the construction of the 3rd power of the recursion operator~\eqref{reckdv} we should
``trivialise'' the following conserved densities of the Korteweg--\/de Vries equation: $u$, $u^2+u_{xx}$ and
$u_{4x} + 6uu_{xx} + 5u_x^2 + 2u^3$.

Let $S$ and $Q$ satisfy the linearized equation for~\eqref{eq5ordF}. The correspondence between fields is the
following $u\mapsto S$, $\xi\mapsto Q$. The linearized system of nonlocalities for the generalisation of those
conservation laws for the supersymmetric equation~\eqref{eq5ordF} is the following:

1) the layer of nonlocalities corresponding to the generalisation of the conserved density~$u$ of the Korteweg--\/de Vries equation
{\small
\begin{align*}
  W_{1;x}={}&S,\\
  W_{1;t}={}& - 15Q u \xi_{x} + S (30u^2 + 10u_{xx} - 15\xi \xi_{x}) - 5Q_{xx}\xi_{x} + Q_{x}(5\xi_{xx} + 15\xi u ) + S_{4x}\\
  {}&{}+ 10S_{xx}u  + 10S_{x}u_{x},
\end{align*}}
2) the layer of nonlocalities corresponding to the generalisation of the conserved density~$u^2+u_{xx}$ of the Korteweg--\/de Vries equation 
{\small
\begin{align*}
  W_{2;x}={}&2S u  + 2Q_{x}\xi,\\
  Q_{2;x}={}&Q u  + S \xi,\\
  W_{2;t}={}&  Q ( - 30u^2\xi_{x} + 30\xi u u_{x}) + S (60u^3 + 40u u_{xx} + 2u_{4x} - 30\xi_{xx}\xi_{x} - 120\xi u \xi_{x}\\
  {}&{} - 20\xi \xi_{xxx}) + 2Q_{5x}\xi  - 2Q_{4x}\xi_{x} + Q_{xxx}(2\xi_{xx} + 20\xi u ) + Q_{xx}( - 30u\xi_{x} - 2\xi_{xxx}\\
  {}&{}+ 30\xi u_{x}) + Q_{x}(30u \xi_{xx} + 2\xi_{4x} + 60\xi u^2 + 10\xi u_{xx}) + 2S_{4x}u  - 2S_{xxx}u_{x}\\
  {}&{}+ S_{xx}(20u^2 + 2u_{xx} - 10\xi \xi_{x}) + S_{x}( - 2u_{xxx} - 30\xi \xi_{xx}),\\
  Q_{2;t}={}&  Q (15u^3 + 10u u_{xx} + u_{4x} + 5u_{x}^2 - 5\xi_{xx}\xi_{x}) + S (20u \xi_{xx} + \xi_{4x} - 5\xi_{x}u_{x} + 45\xi u^2\\
  {}&{} + 10\xi u_{xx}) + Q_{4x}u  - Q_{xxx}u_{x} + Q_{xx}(10u^2 + u_{xx} + 5\xi \xi_{x}) + Q_{x}( - 5u u_{x} - u_{xxx}\\
  {}&{} - 5\xi \xi_{xx}) + S_{4x}\xi - S_{xxx}\xi_{x} + S_{xx}(\xi_{xx} + 10\xi u ) + S_{x}( - 5u \xi_{x} - \xi_{xxx} + 10\xi u_{x}),
\end{align*}}
3) the layer of nonlocalites corresponding to the generalisation of the conserved densities~$u_{4x} + 6uu_{xx} + 5u_x^2 + 2u^3$ of the Korteweg--\/de Vries equation
{\small
\begin{align*}
  W_{3;x}={}& - 9Q_2 \xi u  + 3Q u \xi_{x} + S (6u^2 + 3\xi \xi_{x}) - Q_{xxx}\xi  - 3Q_{x}\xi u  + 2S_{xx}u,\\
  Q_{3;x}={}&Q_2 (6u^2 - 6\xi \xi_{x}) - 6W_2 \xi u  + 14S u \xi_{x} + 2Q_{xxx}u  + 7Q_{x}u^2 - 2S_{xxx}\xi,\\
  W_{3;t}={}& Q_2 ( - 90u^2\xi_{xx} - 9u \xi_{4x} + 45u \xi_{x}u_{x} + 9\xi_{xxx}u_{x} - 9\xi_{xx}u_{xx} + 9\xi_{x}u_{xxx} - 135\xi u^3\\
  {}&{} - 90\xi u u_{xx} - 9\xi u_{4x} - 45\xi u_{x}^2 + 45\xi \xi_{xx}\xi_{x}) + Q (45u^3\xi_{x} + 9u^2\xi_{xxx} + 3u \xi_{5x}\\
  {}&{} - 42u \xi_{xx}u_{x} + 21u \xi_{x}u_{xx} - 3\xi_{4x}u_{x} + 3\xi_{xxx}u_{xx} - 3\xi_{xx}u_{xxx} + 3\xi_{x}u_{4x} + 42\xi_{x}u_{x}^2 \\
  {}&{} - 15\xi u u_{xxx} - 45\xi u_{xx}u_{x})  + S (180u^4 + 300u^2u_{xx} + 32u u_{4x} + 138u \xi_{xx}\xi_{x}\\
  {}&{} - 32u_{xxx}u_{x} + 16u_{xx}^2 + 16\xi_{4x}\xi_{x} - 16\xi_{xxx}\xi_{xx}  + 135\xi u^2\xi_{x} + 78\xi u \xi_{xxx} + 13\xi \xi_{5x}\\
  {}&{} + 93\xi \xi_{xx}u_{x} + 111\xi \xi_{x}u_{xx}) + Q_{6x}\xi_{x} + Q_{5x}( - \xi_{xx} - 13\xi u )
      + Q_{4x}(6u \xi_{x} + \xi_{xxx}\\
  {}&{} - 32\xi u_{x}) + Q_{xxx}( - 26u \xi_{xx} - \xi_{4x} + 26\xi_{x}u_{x} - 24\xi u^2 - 48\xi u_{xx}) + Q_{xx}(69u^2\xi_{x}\\
  {}&{} + 16u \xi_{xxx} - 15\xi_{xx}u_{x} + 17\xi_{x}u_{xx} - 48\xi u u_{x} - 22\xi u_{xxx}) + Q_{x}( - 69u^2\xi_{xx} \\
  {}&{} + 4u \xi_{4x} + 45u \xi_{x}u_{x} - \xi_{xxx}u_{x} - 2\xi_{xx}u_{xx} + 5\xi_{x}u_{xxx} - 45\xi u^3 - 66\xi u u_{xx}\\
  {}&{} - 8\xi u_{4x} - 87\xi u_{x}^2) + 2S_{6x}u  - 2S_{5x}u_{x} + S_{4x}(26u^2 + 2u_{xx} + 8\xi \xi_{x})\\
  {}&{} + S_{xxx}(28u u_{x} - 2u_{xxx} + 22\xi \xi_{xx}) + S_{xx}(120u^3 + 112u u_{xx} + 2u_{4x} - 28u_{x}^2\\
  {}&{} + 22\xi_{xx}\xi_{x} + 81\xi u \xi_{x} + 48\xi \xi_{xxx}) + S_{x}(180u^2u_{x} + 48u u_{xxx} - 16u_{xx}u_{x}\\
  {}&{} + 16\xi_{xxx}\xi_{x} + 3\xi u \xi_{xx} + 32\xi \xi_{4x} + 174\xi \xi_{x}u_{x}) + \xi Q_{7x},
\end{align*}
\begin{align*}
Q_{3;t}={}& Q_2 (90u^4 + 120u^2u_{xx} + 12u u_{4x} - 180u \xi_{xx}\xi_{x} - 12u_{xxx}u_{x} + 6u_{xx}^2 - 12\xi_{4x}\xi_{x}\\
  {}&{} + 12\xi_{xxx}\xi_{xx} - 270\xi u^2\xi_{x} - 60\xi u \xi_{xxx} - 6\xi \xi_{5x} - 90\xi \xi_{xx}u_{x} - 30\xi \xi_{x}u_{xx})\\
  {}&{} + Q (45u^3u_{x} + 24u^2u_{xxx} + 84u u_{xx}u_{x} - 12u \xi_{xxx}\xi_{x} - 28u_{x}^3 - 84\xi_{xx}\xi_{x}u_{x}\\
  {}&{} + 60\xi u^2\xi_{xx} - 24\xi u \xi_{4x} - 30\xi u \xi_{x}u_{x} - 96\xi \xi_{xxx}u_{x} - 84\xi \xi_{xx}u_{xx} - 36\xi \xi_{x}u_{xxx})\\
  {}&{} + W_2 ( - 60u^2\xi_{xx} - 6u \xi_{4x} + 30u \xi_{x}u_{x} + 6\xi_{xxx}u_{x} - 6\xi_{xx}u_{xx} + 6\xi_{x}u_{xxx} - 90\xi u^3\\
  {}&{} - 60\xi u u_{xx} - 6\xi u_{4x} - 30\xi u_{x}^2 + 30\xi \xi_{xx}\xi_{x}) + S (660u^3\xi_{x} + 288u^2\xi_{xxx} + 34u \xi_{5x}\\
  {}&{} - 28u \xi_{xx}u_{x} + 590u \xi_{x}u_{xx} - 34\xi_{4x}u_{x} + 34\xi_{xxx}u_{xx} - 34\xi_{xx}u_{xxx} + 34\xi_{x}u_{4x}\\
  {}&{} + 28\xi_{x}u_{x}^2 + 135\xi u^2u_{x} - 102\xi u u_{xxx} - 20\xi u_{5x} - 366\xi u_{xx}u_{x} - 72\xi \xi_{xxx}\xi_{x})\\
  {}&{} - 2Q_{6x}u_{x} + Q_{5x}(27u^2 + 2u_{xx} - 10\xi \xi_{x}) + Q_{4x}(36u u_{x} - 2u_{xxx} - 20\xi \xi_{xx})\\
  {}&{} + Q_{xxx}(106u^3 + 124u u_{xx} + 2u_{4x} - 36u_{x}^2 - 20\xi_{xx}\xi_{x} + 72\xi u \xi_{x})\\
  {}&{} + Q_{xx}(121u^2u_{x} + 36u u_{xxx}  - 12u_{xx}u_{x} + 10\xi_{xxx}\xi_{x} - 30\xi u \xi_{xx} + 20\xi \xi_{4x}\\
  {}&{} + 144\xi \xi_{x}u_{x}) + Q_{x}(165u^4 + 295u^2u_{xx} + 24u u_{4x}  + 28u u_{x}^2 - 30u \xi_{xx}\xi_{x}\\
  {}&{} - 24u_{xxx}u_{x} + 12u_{xx}^2 + 10\xi_{4x}\xi_{x} - 10\xi_{xxx}\xi_{xx} - 42\xi u \xi_{xxx} + 10\xi \xi_{5x} \\
  {}&{} - 114\xi \xi_{xx}u_{x} + 30\xi \xi_{x}u_{xx}) + 2S_{6x}\xi_{x} + S_{5x}( - 2\xi_{xx} - 20\xi u ) + S_{4x}(44u \xi_{x}\\
  {}&{}+ 2\xi_{xxx} - 80\xi u_{x})  + S_{xxx}(16u \xi_{xx} - 2\xi_{4x} + 36\xi_{x}u_{x} - 36\xi u^2 - 140\xi u_{xx})\\
  {}&{}+ S_{xx}(325u^2\xi_{x} + 104u \xi_{xxx} - 82\xi_{xx}u_{x}  + 94\xi_{x}u_{xx} - 306\xi u u_{x} - 140\xi u_{xxx})\\
  {}&{}+ S_{x}(91u^2\xi_{xx} + 56u \xi_{4x} + 386u \xi_{x}u_{x} - 22\xi_{xxx}u_{x}  - 12\xi_{xx}u_{xx} + 46\xi_{x}u_{xxx}\\
  {}&{}+ 45\xi u^3 - 246\xi u u_{xx} - 80\xi u_{4x} - 414\xi u_{x}^2 - 174\xi \xi_{xx}\xi_{x})  + 2u Q_{7x} - 2\xi S_{7x}. 
\end{align*}}
Here $W_1$, $W_2$, $W_3$ are bosonic fields and $Q_2$, $Q_3$ are fermionic fields.
Let us note that the generalisation of the conservation law of~\eqref{bKdV} with the density
$u_{4x} + 6uu_{xx} + 5u_x^2 + 2u^3$ is no longer a local conservation law for~\eqref{eq5ordF}.

The weights of fields are as follows
\begin{align*}
{}&  |x| = -1,\  |t|=-5,\  |u| = |S| = 2,\  |\xi|= |Q| = \tfrac32,\\ 
{}&  |W_1|= 1,\  |W_2| = 3,\  |W_3|= 5,\  |Q_2| = \tfrac52,\  |Q_3| = \tfrac{11}2.
\end{align*}

Using the technique described above we obtain the following recursion operator (cf.~\cite{TianLiu5ord})
\[
  \cR\left(\begin{pmatrix} S \\ Q  \end{pmatrix}\right) 
  =   \begin{pmatrix} S^\prime \\ Q^\prime  \end{pmatrix},
\]
where
{\small
\begin{align*}
  S^\prime={}& 3Q_3 \xi_{x} + Q_2 ( - 42u \xi_{xx} - 6\xi_{4x} - 42\xi_{x}u_{x} - 18\xi u^2) + Q ( - 51u^2\xi_{x} - 12u \xi_{xxx} \\
  {}&{} + 2\xi_{5x} - 26\xi_{xx}u_{x} - 22\xi_{x}u_{xx} + 18\xi u u_{x} - 2\xi u_{xxx}) + 4W_3 u_{x} + W_2 (24u u_{x} \\
  {}&{} + 4u_{xxx} + 6\xi \xi_{xx}) + W_1 (120u^2u_{x}   + 40u u_{xxx} + 4u_{5x} + 80u_{xx}u_{x} - 20\xi_{xxx}\xi_{x} \\
  {}&{} - 60\xi u \xi_{xx} - 60\xi \xi_{x}u_{x}) + S (128u^3 + 192u u_{xx}  + 24u_{4x} + 144u_{x}^2 - 16\xi_{xx}\xi_{x} \\
  {}&{} - 180\xi u \xi_{x} - 30\xi \xi_{xxx}) - 12Q_{4x}\xi_{x} + Q_{xxx}( - 6\xi_{xx} + 36\xi u )  + Q_{xx}( - 12u \xi_{x}\\
  {}&{} + 8\xi_{xxx} + 58\xi u_{x}) + Q_{x}( - 12u \xi_{xx} + 8\xi_{4x} - 22\xi_{x}u_{x} + 72\xi u^2 + 30\xi u_{xx})  + 2S_{6x} \\
  {}&{} + 24S_{4x}u + 60S_{xxx}u_{x} + S_{xx}(96u^2 + 80u_{xx} - 30\xi \xi_{x}) \\
  {}&{} + S_{x}(280u u_{x} + 60u_{xxx} - 54\xi \xi_{xx}),
\end{align*}
\begin{align*}
  Q^\prime ={}& 3Q_3 u  + Q_2 (42u u_{x} + 6u_{xxx}) + Q (35u^3 + 48u u_{xx} + 2u_{4x} + 40u_{x}^2 - 6\xi_{xx}\xi_{x}\\
  {}&{} + 4\xi \xi_{xxx}) + 4W_3 \xi_{x}  + W_2 ( - 12u \xi_{x} - 2\xi_{xxx} - 6\xi u_{x}) + W_1 (60u^2\xi_{x} + 40u \xi_{xxx}\\
  {}&{} + 4\xi_{5x} + 60\xi_{xx}u_{x} + 20\xi_{x}u_{xx} + 60\xi u u_{x})   + S (114u \xi_{xx} + 22\xi_{4x} + 104\xi_{x}u_{x}\\
  {}&{} + 72\xi u^2 + 30\xi u_{xx}) + 2Q_{6x} + 24Q_{4x}u  + 48Q_{xxx}u_{x}   + Q_{xx}(54u^2 + 38u_{xx} + 4\xi \xi_{x})\\
  {}&{} + Q_{x}(126u u_{x} + 14u_{xxx}) + 12S_{xxx}\xi_{x} + S_{xx}(42\xi_{xx} + 42\xi u )\\
  {}&{} + S_{x}(100u \xi_{x} + 46\xi_{xxx} + 54\xi u_{x}).
\end{align*}}

In~\cite{GradNonRem} this system of nonlocalities was used to construct a zero-curvature representation
of~\eqref{eq5ordF} to prove its integrability.

\end{Exmpl}     

\subsection*{Acknowledgements}
This work was supported in part by an NSERC grant to T.\,Wolf who is
thanked by A.\,V.\,K.\ for warm hospitality.
% This work was supported in part by Brock university.
% A.\,V.\,K.\ thanks T.\,Wolf for warm hospitality.
This research was done in part while A.~O.~K. was visiting at New York University Abu Dhabi; the hospitality and
warm atmosphere of this institution are gratefully acknowledged.

% relax definitions

\let\Lin\relax
\let\sym\relax
\let\id\relax

\let\sD\relax
\let\BBN\relax
\let\BBC\relax
\let\cR\relax

\end{document}